%% file: paper1.tex
\documentstyle[12pt,aasms]{article}
\begin{document}
\def\as1{$A_1$}
\def\dv{$|\Delta v|$}
\def\Kp{K$^\prime$\ }
\def\a1ave{$\langle A_1 \rangle$}
\def\db{$\Delta$B\ }

\def\etal{{\it et al.~}}
\def\mn{\noindent}
\def\apj{\noindent}
\def\nature{\noindent}
\def\refbook{\noindent}

\input{macro.tex}
\title{More Satellites of Spiral Galaxies}
\author{Dennis Zaritsky$^1$, Rodney Smith$^2$, Carlos Frenk$^3$, and 
Simon D.M. White$^4$}

\vskip 1in

\noindent
$^1$ Carnegie Observatories, 813 Santa Barbara St., Pasadena,
CA, 91101 and UCO/Lick Observatory and Board of Astronomy and Astrophysics,
Univ. of California, Santa Cruz, CA, 95064, E-Mail: dennis@ucolick.org

\noindent
$^2$ Department of Physics and Astronomy, P.O. Box 913, University of Wales, 
College of Cardiff, Cardiff CF2 3YB, Wales, E-Mail: R.Smith@astro.cf.ac.uk

\noindent
$^3$ Department of Physics, South Road, University of Durham, Durham,
DH1 3LE, England, E-Mail: c.s.frenk@durham.ac.uk

\noindent
$^4$ Max-Planck-Institut f\"ur Astrophysik, Karl-Schwarzschild-Strasse
1, D-85748 Garching bei M\"unchen, E-mail: swhite@mpa-garching.mpg.de
\vskip 1in

\abstract{We present a revised and expanded catalog of satellite
galaxies of a set of isolated spiral galaxies similar in luminosity
to the Milky Way. This sample of 115 satellites, 
69 of which were discovered in our multifiber redshift survey,
is used to further probe the
results obtained from the original sample (Zaritsky \etal 1993). 
The satellites are, by definition,
at projected separations $\ltsim$ 500 kpc, have absolute recessional 
velocity differences with respect to the parent spiral of less than
500 km s$^{-1}$ and are at least 2.2 mag fainter than their associated
primary galaxy. A key characteristic of this survey is the
strict isolation of these systems, which simplifies any dynamical
analysis. 

We find no evidence for a decrease in the velocity
dispersion of the satellite system
as a function of radius out to galactocentric radii of 400 kpc, 
suggesting that the halo extends
well beyond 200 kpc. Furthermore, the new sample affirms our previous 
conclusions (Zaritsky
\etal 1993) that (1) the velocity
difference between a satellite and its primary is not strongly correlated
with the rotation speed of the primary, (2) the system of satellites
has a slight net rotation ($34 \pm 14$ km s$^{-1}$) in the
same sense as the primary's disk, and (3) 
that the halo mass of an $\sim L^*$ spiral galaxy 
is in excess of $2 \times 10^{12} M_\odot$. 

}

\section{Introduction}

Despite
the more than 60 years since Zwicky (1933) and Smith's (1936) measurements
of the velocity dispersions in the Coma and Virgo galaxy clusters,
and the roughly 20 years since the extended neutral hydrogen galaxy 
rotation curves of Rogstad and Shostak (e.g. 1971), Bosma (1978) 
and Rubin, Thonnard, and Ford (1978), we know little about the distribution
of the more than 90\% of the mass of Universe that consists of dark
matter. As we have 
discussed previously in related work (Zaritsky \etal 1993
and Zaritsky and White 1994; hereafter ZSFW and ZW respectively), 
satellite galaxies are currently the best available 
probes of the distribution of mass in galaxy halos
at large radii ($>$ 100 kpc) 
from individual galaxies. 

In our previous
studies, we presented results based on our initial sample of 
69 satellites of 45 primaries. In particular, we concluded
that the value of the typical halo mass enclosed within a
200 kpc radius lies between $1.5\times
10^{12}$ and $2.6\times 10^{12} M_\odot$ (90\% confidence limits; ZW). 
However, those studies
left several open questions: (1) does
the asymmetric distribution of satellite-primary velocity 
differences (Arp 1982; Arp \& Sulentic 1985; Zaritsky 1992)
persist in ever-increasing samples and is it indicative
of an interloper fraction much larger than the estimated 0.1
(Zaritsky 1992),
(2) will the apparent lack of a correlation between the
satellite-primary velocity difference and the primary disk rotation 
speed persist with increased statistics and
a larger range of rotation speeds, (3) is the slight net prograde
rotation (29 $\pm$ 21 km s$^{-1}$ relative to the primary's disk rotation;
ZSFW) 
of the satellite system real, and
(4) what is the halo mass distribution beyond 200 kpc?

In this paper, we address these questions with our revised and 
expanded sample of satellite galaxies. We continue to
overcome the limitation presented by the small number of bright
satellites around
any particular galaxy through a statistical treatment of satellites
around a large, well-specified, sample of isolated spiral
galaxies. Our fundamental
assumption is that an ensemble of satellites belonging
to a carefully selected sample of primary galaxies can be treated
as if the satellites all belong to a single ``typical" primary galaxy. The 
validity of this assumption depends on the correspondence between
the directly observable characteristics of spiral galaxies (such
as luminosity) and the 
indirectly observable characteristics of their dark matter halos (such
as mass). 
The ensemble approach is only viable {\it either} if there is no
correspondence between disk and halo properties or if the correspondence
is known.
Our previous results suggest that there is little or no 
correspondence between the mass of the 
central galaxy and its halo (ZW), thereby supporting
our fundamental 
assumption to even a greater degree than one might naively expect.
Nevertheless, one of the principal aims 
of this study is to further test this assumption.

We present the new satellite sample and discuss some basic 
observational results. A  detailed analysis of the dynamics
of the satellite system will be presented elsewhere.
We discuss the criteria by which we select primaries and
satellites, and the details of how we compile the sample in \S2.
We address the implications of the new sample on the questions
posed above in \S3.

\section{The Data}

\subsection{Primary and Satellite Selection}

The sample consists of data collected by the authors over the last
eight years at a variety of telescopes and from
the literature. Satellite identification and
follow-up velocity measurements are described by ZSFW for those
satellites included in the first sample. The new satellite identifications
are from multifiber redshift surveys at either the Las Campanas
Observatories (LCO) or the Anglo-Australian Telescope (AAT). 

Once again, we attempt to define a homogeneous sample of isolated
primaries. We have slightly modified our criteria from those adopted by
ZSFW in order 
to expand the sample and to test whether some of the previous 
conclusions depend on the restrictiveness of the selection criteria.
Principally, we expand the allowed magnitude range of primaries to examine
whether the observed lack of a correlation between 
the absolute value of the satellite-primary velocity difference, \dv, and
the disk rotation speed for the ZSFW sample
is due to the smaller range of primary luminosities in that sample. 
The primaries in the current sample span $-22.4 < M_B < -18.8$ 
(for H$_0 = 75$ km s$^{-1}$ Mpc$^{-1}$ as adopted throughout), although
that range is not uniformly sampled (cf. Figure 1). We also
relax the criteria against barred primary galaxies. The exclusion of barred
galaxies was motivated in ZSFW by the possibility that barred galaxies might 
reside in halos which are insufficiently massive to suppress bar
formation (and so significantly different from the halos of
unbarred galaxies). However, the selection of unbarred galaxies 
is ill-defined because many galaxies are mixed type
and close inspection of ``unbarred" galaxies can reveal the
presence of a small bar (\eg M51: Zaritsky, Rix, \& Rieke 1993). 
We maintain nearly the same range of allowed recessional velocities
as in ZSFW, 1000 to 7500\kms. Finally, the isolation criteria 
remain the same as in ZSFW $-$ all companions
within 1000\kms\ in velocity and 500 kpc projected separation must be at least
2.2 mag fainter than the primary and those within 1000\kms\ and 
between 500 kpc and 1 Mpc projected separation must be at least
0.7 mag fainter.

Our principal source of candidate primaries is the 
CfA redshift survey catalog
(Huchra 1987). A follow-up search for companions and nearby galaxies
is done using the NASA Extragalactic Database (NED). Magnitudes for
possible nearby bright neighbors are estimated, when not available
from NED, using the Space Telescope Digitized Sky Survey
in the same manner in which satellite magnitudes are measured (see
\S 2.2 for details). The introduction of quantitative, consistent, and
reproducible magnitude measurements is a significant improvement
over what was available for ZSFW. This new information has led
to the rejection of some fields that were included in ZSFW. In 
Table 1 we present our justifications for the rejection of certain fields 
and satellites from the ZSFW sample. In total, 
13 satellites from the previous sample are not included in the
current sample, principally as a result of the isolation criteria
and new magnitude measurements. 

Once a sample of suitable primaries is identified, 
candidate satellite galaxies and their coordinates
are obtained either through 
visual inspection of Palomar or ESO sky survey plates (for
the LCO observations) or of APM identified galaxies
(for the AAT observations). The selection
is susceptible to biases, although
it is independent of the most important characteristic of the 
satellite to us, its recessional velocity relative to the
primary. Typically, the satellites
are brighter than $m_B = 18.5$ and have angular sizes greater than ten arcsec. 
No conscious selection (\eg surface brightness, morphology, or
nearness to primary) is introduced. The fiber assignment can introduce
some biases (\eg two candidate satellites separated by a small angular
distances cannot both be observed in a single fiber setup), 
as can the telescope/spectrograph
combination (\eg outer field vignetting
decreases the likelihood of observing objects at the edges of the 1.5$^{\circ}$
diameter field of view). Therefore, as stressed by ZW, a correct dynamical
analysis of these data will examine the velocity distribution
at each projected radius, rather than relying on a statistic that 
incorporates the radial satellite 
distribution (\eg the projected mass estimator, $r_p\Delta
v^2$).

With the candidate satellites in hand, the next task is to
separate satellites from pretenders by
measuring recessional velocities.
The LCO satellite redshift survey was done with the multifiber
spectrograph and 2D-Frutti detector (Shectman \etal 1992) 
on the du Pont 2.5m telescope. The fiber apertures are 3 arcsec
in diameter and the fibers are placed manually into a plug plate. 
A 1200 line mm$^{-1}$ grating 
provides wavelength coverage from $\sim$ 3600 to 5300 \AA\ 
with a resolution of approximately 2.5 \AA. 
The observations occurred during runs in 1992 March,
1993 February, and 1993 November. We typically observed 30
to 40 galaxies per field for two hours 
in each of 45 fields (combined for the three runs).

The data were reduced using IRAF\footnote{IRAF is distributed by the
National Optical Astronomy Observatories, which are operated
by AURA, Inc., under contract to the NSF.}. Velocities
are obtained from either the wavelength of bright emission
lines or from a cross-correlation analysis (cf. Tonry \& Davis 1979).
The velocity uncertainty in a single measurement,
$\sim$ 70\kms, is too large for
our dynamical analysis, but is sufficiently precise for the identification
of satellites. The fraction of targeted objects for which we measure
a redshift is 0.78.
Follow-up long-slit observations are obtained
to confirm the satellites and measure precise velocities. 

Long-slit observations of satellites identified from the multifiber survey
and of other satellites that require follow-up
observations (\ie satellites found in the literature with
large velocity uncertainties or 
very faint satellites in ZSFW with dubious velocity
measurements) were taken at the 2.5 du Pont telescope
at LCO using the Modular Spectrograph.
A 1200 line mm$^{-1}$ grating provides wavelength
coverage from about 4650 to 5900 \AA\ with a resolution of about
3 \AA. Although the resolution is comparable to that of the
fiber spectra, the stability and linearity of the wavelength
solution results in velocities that have $1\sigma$ uncertainties
of $\le$ 20\kms\ (see below). These 
long-slit observations were taken in 1993 February,
1993 November, 1994 February, and 1994 November. 
In addition to observing the satellites, 
we observed most of the primary galaxies, with the slit placed
along the major axis as seen on the acquisition camera, to measure the 
direction of disk rotation. 
The observation of primary galaxies
also serves as an external test of our velocity measurements 
(neutral hydrogen measurements are available for
most of the primaries from the compilation by Huchtmeier and Richter (1989)).

In Figure 2, we combine primary and satellite
velocities for which we have velocities from both the literature
and long-slit observations from LCO, and we plot one-half the difference
between the two values for each galaxy. The 
standard deviation from a value of zero is 15\kms.
We overplot a Gaussian of $\sigma = 15$\kms\ for comparison.
Accepted values from the literature typically have quoted
uncertainties that are $\ltsim 10$\kms.
Because \dv\ is the difference between the primary
and satellite velocities, the uncertainty in \dv\ is
$\sqrt 2 \times \sigma = 21$\kms. We conservatively adopt a \dv\ 
$1\sigma$ uncertainty of 30\kms.  

The AAT observations were taken during May 1993 using the
AUTOFIB automatic fiber positioner feeding the RGO Spectrograph. The
60 fibers each have an angular aperture of 2.1 arcsec. A 1200 line mm$^{-1}$
grating was used with the 25 cm camera and a Tektronix CCD to give
a resolution of $\sim$ 2 \AA\ and a wavelength range of $\sim$
4700 to 5500 \AA. As many of the fibers as possible were placed on
galaxies such that there was no fixed limiting apparent magnitude. Each
field was observed for a total of 8000 sec, split into four
2000 sec exposures.

The AAT data were reduced using the FIGARO spectroscopic data reduction
package of the STARLINK suite of astronomical software. Typical velocity
errors for spectra with only absorption lines is 20 km s$^{-1}$, so no
follow-up long-slit observations were necessary. 
Fiber apertures were placed 
along the major axis of the central galaxy for a few targets,
but generally only the redshift of the core
of the primary was observed. Unfortunately, the seeing was generally
poor, often greater than 3 arcsec, so the success rate at
determining redshifts was not as high as for the LCO observations. 

The complete sample of satellite galaxies is presented in Table 2.
In Column (1) we list the name of the primary (N stands for NGC, 
I for IC, and A for Anonymous) or the letter designation of the
satellite. In Columns (2) and (3) we list the right ascension and
declination of the object in 1950.0 epoch coordinates. In Column
(4) we present the blue absolute magnitude of the galaxy as given by 
the apparent magnitude (either from NED,
from our own measurements as described below, or from the average of the
two when both are available) and the distance as calculated using
the primary galaxy's recessional velocity, the simple Virgocentric
infall model discussed in ZSFW, and $H_0= 75$\kms\ Mpc$^{-1}$. 
In Column (5) we list the galaxy's recessional
velocity. The recessional velocities for the
primaries are from the average of neutral hydrogen observations when
available, or from our own observations, and lastly, if neither
of those is available (one case, NGC 6948) from the literature at 
large. The last column (Column 13), which presents the notes, identifies
the velocity sources. In Column (6) we list the projected separation
in kpc between the satellite and primary. In Column (7) we present
our measurement of the galaxy's semimajor axis length (in kpc).
In Column (8) we list the average value of the width of the
neutral hydrogen profile at 20\% of peak
intensity level in\kms\ from the compilation of Huchtmeier and Richter 
(1989). In Column (9) we list the inclination of the galaxy to the
line-of-sight. The inclination values come from Huchtmeier 
and Richter for the primaries and
are estimated from the
axis ratio as measured from the sky survey plates (see below)
for the satellites.
In Column (10) we present the angle between the radius vector
from the primary to the satellite and the primary's
major axis. In Column (11) we
identify whether the satellite has observable emission lines in its
spectrum (check means yes, blank means no, ellipses means no
spectrum available). 
In Column (12) we identify whether the satellite is on a
prograde or retrograde orbit relative to the rotation of the 
primary's disk.
Finally, in Column (13) we provide notes on the sources of the
recessional velocity data. The code is as follows: H for Huchtmeier
and Richter (1989), M for our observations at the MMT telescope 
(described in ZSFW), N for data from NED, C for our observations
at CTIO (described in ZSFW), LC for our long-slit observations
at LCO, LCf for our fiber observations at LCO, MX for our observations
using the MX Spectrograph at the Steward 2.3m telescope (described
in ZSFW), W for our observations using the William Herschel telescope
(described in ZSFW), and A for our observations using the AAT.

\subsection{Supplementary Data}

The sizes and magnitudes of the satellites, and their position angles
relative to the primary's major axis 
(cf. Table 2) are measured from the Space
Telescope Digitized Sky Survey images. Sizes are measured interactively
off the monitor at the faintest surface brightness
visible. The magnitudes are measured using the aperture photometry (PHOT)
task in IRAF. The aperture is interactively set for each object to the 
smallest radius at which the 
magnitude converges. The photometric calibration 
was done independently for POSS, UK-Schmidt, and 2nd generation POSS
survey plates using satellite 
magnitudes available in the literature.
The dispersion about all three calibrations was 
$\ltsim 0.5$ mag, so we conservatively adopt 0.5 mag
as our magnitude uncertainty. Although
this approach clearly has pitfalls (\eg photographic plate non-linear
photometry solutions, plate-to-plate variations, poor spatial sampling,
and unknown color terms),
it is a more quantifiable and robust method than the visual magnitude
estimation done by ZSFW. In fact, this new method led to 
significant revisions in the magnitudes of a few objects, which
led to their rejection from the sample (see \S 2.1 and Table 1). 

The magnitude distribution of satellite galaxies is shown in Figure 3.
As in ZSFW, the satellites principally have $M_B > -18$ and 
$3 < \Delta M_B < 5.5$
(Figure 4), suggesting (for the same mass-to-light ratio
as their parents) that they are generally between 1/10th and 1/150th
as massive as their parents (the mode of the distribution
corresponds to a satellite-primary
mass ratio of 1:30). As evident in Figure 4, 
there is no systematic difference
in the luminosity distribution of the fiber identified satellites and those
identified from the literature. 

Average surface brightnesses are estimated using the major and minor
axis lengths of the satellites and their magnitudes. The surface brightness
can be calculated assuming either that the isophotal ellipticities reflect 
true asymmetries in the galaxy (and so no inclination correction
is included) or that the ellipticities are directly due to inclination
and that the galaxy is optically thin. We show the relationship
between surface brightness and projected separation in Figure 5 for
both treatments. The correlation between surface brightness
and projected separations is statistically insignificant in
both panels.

Our determination of whether an object has emission lines is based
on a visual examination of the observed spectra. We 
possibly miss emission lines due to the placement
of the fiber or slit on the object, 
or due to the low signal-to-noise of the lines; 
however, there is no bias that will produce spurious
emission lines. Of the satellites for which we obtained spectra, 
86\% have detectable emission lines. 

The orbital orientation of a satellite is determined for the 57 systems
for which the direction of the primary's spin has been measured. The
satellite is on a prograde, ``p" orbit if its recessional velocity is
in the same sense as the side of the disk nearest to it,
and retrograde, ``r", otherwise. 

\section{Discussion}

\subsection{``Numerology"}

The new sample of satellites consists of 115 satellites around 69
primaries. These are divided almost equally between satellites
presented in ZSFW and new satellites (56 vs. 59, respectively), and
there are about 1.5 times as many fiber satellites as literature
satellites (69 vs. 46, respectively; where satellites ``rediscovered"
in the multifiber surveys are categorized as fiber satellites).

As shown in Figure 6, systems range from very rich (with 4 or 5 members),
which are suspiciously like galaxy groups, to single satellite
members. The groups are a particularly important class because they
have the potential to skew statistics in a correlated manner.
For example, if the five satellites of NGC 1961 are in fact a group
projected near NGC 1961, then they could conceivably {\it all}
have large projected separations and velocity differences. Physically,
there may also be subtle dynamical interactions between the
satellites themselves, which are missed when satellites
are treated as independent particles. 
The average number of satellites per primary, for 
primaries with detected satellites, is 1.7. This average is slightly
larger than the 1.5 average for the ZSFW sample. In both cases, this
average represents a lower limit to the number of satellites because
the surveys are incomplete.

The decline in the number of systems
with satellite richness (Figure 6) is quite regular. If we treat the 
presence of a satellite as a random event with probability, $p$,
then the probability  of finding a system with $n$ satellites
is $p^n$. We fit this expression to the numbers
of systems with $n = 1,5$ to find the effective number of
fields observed and $p$. 
The best fit value for the probability of
finding a satellite around an
isolated spiral galaxy and within the magnitude range we probe
is $0.43 \pm 0.06$ and the fit to the data has
$\chi^2 = 0.38$. We also find that the effective number of fields
searched is 94 $\pm$ 25, which implies that the average number
of satellites per field spiral galaxies is 1.2 $\pm$ 0.3.
We overplot this model in Figure 6. 
The quality of the fit
supports the assumption that the satellites are independent
(\eg they do not preferentially come in pairs), which simplifies
any analysis and justifies treating each satellite as an independent
datum, and that the typical galaxy has at least one satellite, which
implies that spiral galaxies with satellites are not an atypical subset of
spirals.  

\subsection{The Asymmetry of the Satellite Velocity Distribution}

The distribution of satellite
galaxy velocities, and to a lesser extent binary galaxy velocities,
are skewed so that a satellite, or the fainter of two
binary galaxies, has a greater than 50\% probability of having
a larger recessional velocity than the primary galaxy (Arp 1982; Arp
\& Sulentic 1985;
Zaritsky 1992). The 
naive expectation, when dealing with a system of presumably bound 
objects, is that satellite recessional velocities will be divided equally
into approaching and receding objects. The 
asymmetry (present in existing samples
at a level of about 59:41, receding to approaching
satellites for the projected separation criteria of this study; Zaritsky 1992)
has been interpreted by some (Arp 1982; Arp \& Sulentic 1985)
as an indication of a non-Doppler component of the redshift.
Zaritsky (1992) presented an analysis of the existing satellite samples
and concluded that observational biases in the selection of satellite 
galaxies naturally leads to a slight velocity asymmetry
and that the observed asymmetry was 
consistent with those biases plus counting statistic fluctuations
in the ZSFW sample. However,
if the asymmetry persists at the same level in ever increasing
samples, the effect would become more and more difficult to 
explain in terms of selection biases and counting statistics.

The current sample still has an excess of positive $\Delta v$ satellites, 
although at a lower level than the ZSFW sample. If we quantify the 
asymmetry by the ratio of the number of satellites with
positive $\Delta v$ to the total number of satellites, $P/T$,
then the current sample has $P/T = 0.57$, relative to 
$P/T = 0.59$ for the ZSFW sample. 
Treating the satellites as purely independent trials and
assuming equal probabilities for negative and positive $\Delta v$, 
the probability
of this asymmetry in a sample of 115 satellites is 0.07. 
ZSFW noted that most of the asymmetry in their sample appears to 
come from the literature sample rather than the fiber sample. 
Dividing our sample into literature and fiber samples illustrates that
that impression 
is borne out by the current sample. The literature sample has $P/T =
0.61$, while the fiber sample has $P/T = 0.54$.
The former has only a probability of 0.09 of occurring, while the latter
has a probability of 0.28. The fiber sample is therefore entirely consistent
with a random distribution (and even more consistent with the expected
distribution, which should be slightly asymmetric 
due to observation selection effects
(Zaritsky 1992)). 

Because unknown physical phenomena, which might cause the 
observed effect, would
affect the literature and fiber samples equally, we assert that
selection biases are the dominant cause of the asymmetric
$\Delta v$ distribution. The
literature sample must be skewed 
because it contains more correlated systems (i.e. projected
groups), for which binomial
statistics underestimate the probability of the observed asymmetries. 
We conclude, because the level of asymmetry has declined
slightly as the sample has grown and because the fiber sample appears
to be free of significant contamination, that the total sample
is consistent with the previous estimate of the interloper
fraction in this type of sample ($\sim$ 0.1). As we discuss next,
other properties of the satellite ensemble further support
our conclusion that interlopers are not a dominant fraction of the
sample.

\subsection{The Disk-Halo Connection?}

ZSFW noted that \dv\ and the primary's disk rotation velocity appear
to be uncorrelated
(in fact there is a slight anti-correlation in the ZSFW sample).
An analogous lack of  correlation was also seen in samples
of binary galaxies (White \etal 1983; Charlton \& Salpeter 1991).
The infall model of ZW allowed for 
the possibility that halo and disk masses may not
track each other, but no 
physical motivation for that behavior was presented. Since then,  
numerical simulations of the process of galaxy formation have suggested
that a strong correlation should not be expected between the
two quantities (Navarro, Frenk, \& White 1996). 

In Figure 7 we 
plot \dv\ vs. 1/2 
the inclination corrected value of the H I profile width for
the associated primary (the amplitude of the rotation curve), $0.5W_i$. 
There is no apparent correlation
between the two quantities, regardless of whether the objects with
\dv\ $>$ 250\kms\ are treated as interlopers or not. At a low
significance level, there may be an indication that 
galaxies with $0.5W_i < 200$\kms\ have satellites with smaller than 
average \dv's.
Given the recent results from simulations on the behavior of 
disks and halos, the lack of a strong signal is not as surprising as
it once was and
may be consistent with the parameterization of halo
rotation curves found
by Navarro \etal (1996). 

The second aspect of the disk-halo connection that we examine
is the net prograde
rotation of the satellite system found by ZSFW. Their measurement
of the net rotation
was just slightly more than 1$\sigma$ different from zero net
rotation (29 $\pm$ 21\kms). The new value of the net rotation velocity,
34 $\pm\ 14$\kms, confirms and
strengthens the previous observation
of a slight net prograde rotation of the satellite systems. 

The histogram of velocities is shown in Figure 8. As in the smaller
ZSFW sample, the peak in the distribution appears at small
retrograde velocities and the distribution has an asymmetric
tail to large prograde values. This asymmetry 
is an entirely different one
from that in $\Delta v$. Because any
interlopers should be evenly distributed
between prograde and retrograde velocities, this asymmetry suggests
that at least some objects with large \dv, which are typically
assumed to be interlopers, may be associated with their
respective primary galaxy.  

Lastly, the 
azimuthal distribution of satellites with respect to the disk plane
is elongated perpendicular to the disk plane at radii $>$ 300
(Zaritsky \etal 1996) and suggests that interlopers are not the dominant
population at large radii. Therefore, both the spatial distribution
and the connection between velocities and disk spin suggest that
some of the systems that one might suspect to be interlopers
are physically associated with the primary galaxy.

\subsection{Radial Velocity vs. Projected Separation}

As suggested in \S3.2 and 3.3, the sample is not strongly
contaminated by interlopers and so should provide a reliable probe
of outer halo dynamics. In Figure 9 we plot the distribution
of $\Delta v$ vs. projected separation, $r_p$. The large majority (100 of 115)
of satellites have $\Delta v$'s between $\pm$200\kms\ indicating
that the ``satellites" are indeed associated with (but not necessary 
gravitationally bound to) the primary. The general behavior of the distribution
is identical to that of the ZSFW sample. The
most noticeable change is the increased sampling of the distribution
at radii $\sim$
400 kpc. 

The influence of the dark matter halo on the satellite
velocities is evident in Figure 10, 
where we have plotted \dv\ 
vs. $r_p$. There is no apparent decline in the velocity
dispersion of satellites as a function of radius. Visual comparison
with the models shown in Figure 2 of White and Zaritsky (1992)
suggests that halos with mass profiles characteristic of
$\Omega_0 < 0.3$ models will be strongly ruled out.
The median \dv\ of satellites at projected
separations $>$ 243 kpc, a radial range chosen for direct
comparison to ZSFW, is 78 \kms, slightly
larger than the value of 67 \kms\ obtained for the ZSFW sample. This 
agreement suggests that the ZSFW and ZW results regarding the mass of halos 
are unlikely to change dramatically once this new sample is carefully
analyzed.

\section{Conclusions}

We have constructed a revised and enlarged sample of satellite
galaxies around isolated spiral galaxies. With this sample we
have addressed several questions regarding the nature of galaxy
halos. In particular, we find that

\noindent
(1) the probability of finding a satellite around these
isolates spiral primaries, for systems having from 1 to 5
satellites, is described by $P = 0.4^n$, where $n$ is the number
of satellites (the probability of not finding a satellite in a system
is then 25\%),

\noindent
(2) the new sample has a less asymmetric $\Delta v$ distribution
than the ZSFW sample, and more importantly, the satellites
identified from our fiber survey, for which the selection criteria
and biases are best known, show no significant $\Delta v$ asymmetry,

\noindent
(3) the primary's disk rotation speed and \dv\ still show no
significant correlation,

\noindent
(4) the net prograde rotation of the satellite system with respect to 
the primary disks is confirmed and is now measured to be
$34 \pm 14$\kms, and

\noindent
(5) the lack of any clear dependence between the
velocity dispersion of the satellite system and projected radius
is even more striking than in the ZSFW sample, thereby supporting
the previous conclusion of large and massive galaxy halos.

The revised and expanded satellite sample presented here
has illuminated three key reasons for
believing that most of the identified satellite galaxies 
are indeed related to their primaries (1) the high concentration
of galaxies at \dv\ $< 200$\kms\ (87\% of the sample), (2)
the net prograde rotation of the system relative to the disk
rotation (distinct from zero rotation
at the 98.5\% confidence level), and (3) the asymmetric 
azimuthal distribution of satellites, with the major axis of the
distribution at large radii aligned with the minor axis of the disk (Zaritsky
\etal 1996). This new
sample confirms the result of large and massive dark
matter halos around isolated spiral galaxies (ZSFW and ZW),
but it also begins to provide detailed information about the
distribution and orbits of outer halo material. Because
of the long dynamical timescale at large radius, the behavior
of this material may provide our best
probe of the initial characteristics of galaxy halos. 

\bigskip
\noindent
Acknowledgments: DZ acknowledges partial financial support 
from the California Space Institute and from NASA through
HF-1027.01-91A, from STScI, which is operated by AURA, Inc., under
NASA contract NAS5-26555. 

\vfill\eject
\noindent
{\bf References}
\bigskip

\apj{Arp, H. 1982}{256}{54}
\apj{Arp, H., \& Sulentic, J.W. 1985}{291}{88}
\refbook{Bosma, A. 1978, Ph.D. Dissertation, Univ. of Groningen}
\apj{Charlton, J.C., \& Salpeter, E.E. 1991}{375}{517}
\refbook{Huchra, J. P. 1987, {\it The CfA Redshift Catalogue}}
\refbook{Huchtmeier, W. K., and Richter, O.-G. 1989, {\it H I Observations
of Galaxies}, (New York: Springer-Verlag) (HR)}
\refbook{Ingerson, T. E. 1987, in {\it Instrumentation for Ground-Based
Optical Astronomy}, ed. L.B. Robinson (New York: Springer-Verlag), p. 222}
\refbook{Ingerson, T. E. 1988 in {\it A.S.P Conference Series Vol. 3,
Fiber Optics in Astronomy}, ed. S.C. Barden (Provo: Astronomical
Society of the Pacific), p. 99}
\mn{Lorrimer, S. J., Frenk, C. S., Smith, R. M., White, S. D. M., and Zaritsky, D. 1994}{269}{696}
\refbook{Navarro, J.F., Frenk, C.S., \& White, S.D.M 1996, preprint}
\aa{Rogstad, D.H., \& Shostak, G.S. 1971}{13}{99}
\apjlett{Rubin, V.C., Ford., W. K., Jr., and Thonnard, N. 1978}{225}{L107}
\pasp{Schmidt, G., Weymann, R. J., and Foltz, C. B. 1989}{101}{713}
\refbook{Shectman, S.A., Schechter, P.L., Oemler, A.A., Tucker, D.,
Kirshner, R.P., \& Lin, H. 1992, in  Clusters and Superclusters of
Galaxies (ed. Fabian, A.C.), (Kluwer: Dordrecht), p. 351}
\apj{Smith, S. 1936}{83}{23}
\aj{Tonry, J., and Davis, M. 1979}{84}{1511}
\mn{White, S.D.M. 1981}{195}{1037}
\mn{White, S.D.M., Huchra, J., Latham, D., \& Davis, M. 1983}{203}{701}
\apj{White, S.D.M., \& Zaritsky, D. 1992}{394}{1}
\apj{Zaritsky, D., 1992}{400}{74}
\nature{Zaritsky, D., Rix, H.-W., \& Rieke, M. 1993}{364}{313}
\apj{Zaritsky, D., Smith, R., Frenk, C. S., \& White, S. D. M., 1993}{405}{464}
\apj{Zaritsky, D., \& White, S.D.M. 1994}{435}{599}
\refbook{Zwicky, F. 1993, {\it Helv. Phys. Acta}, 6, 110}

\clearpage

\begin{table}
\caption{Cause for Omission of ZSFW Data}
\input table1.tex
\end{table}
\clearpage

\setcounter{page}{31}
\centerline{\bf Figure Captions}
\bigskip

\noindent
1 $-$ The distribution of blue absolute magnitudes, $M_B$, 
for the primary galaxies. 

\noindent
2 $-$ Velocity measurement reproducibility. We plot the difference
between Las Campanas long-slit measurements and the average of
long-slit and literature values for both primary and satellite
galaxies observed at Las Campanas. We overlay a Gaussian with $\sigma
= 15$\kms\ and mean $=$ 0 for comparison (the Gaussian is not a fit
to the data).

\noindent
3 $-$ The distribution of blue absolute magnitudes for the satellite
galaxies. 
The unshaded histogram represents the entire sample, while the shaded
one represents only those satellites discovered from the fiber
surveys. The location of Local Group galaxies are added at the
top for comparison.

\noindent
4 $-$ The distribution of primary-satellite magnitude differences.
The unshaded histogram represents the entire sample, while the shaded
one represents only those satellites discovered with the fiber
surveys.

\noindent
5 $-$ The distribution of satellite average surface brightness
versus projected separation. The top panel
contains average surface brightnesses over the projected area of
the satellite ($\pi ab$, where $a$ and $b$ are the satellites semimajor
and semiminor axis). The bottom panel contains average surface
brightnesses over the deprojected area ($\pi a^2$).

\noindent
6 $-$ The distribution of the systems as a function of satellite
number. The connected dots represent the best fit
model assuming that the presence of a satellite is a random variable
(probability of finding $n$ satellites around a field spiral galaxy
is proportional to 0.43$^n$).

\noindent
7 $-$ \dv\ vs 0.5 times the inclination corrected neutral hydrogen
profile width, $0.5W_i$.

\noindent
8 $-$ Distribution of satellite radial velocities relative to the disk.
Positive velocities represent prograde satellites, while
negative ones represent retrograde satellites. 

\noindent
9 $-$ The distribution of $\Delta v$ vs $r_p$ for the satellite 
galaxies. Filled circles represent satellites from the fiber 
surveys and crosses represent those found from literature surveys.

\noindent
10 $-$ The distribution of \dv\ vs. $r_p$ for the satellite
galaxies.

\begin{figure}
\plotone{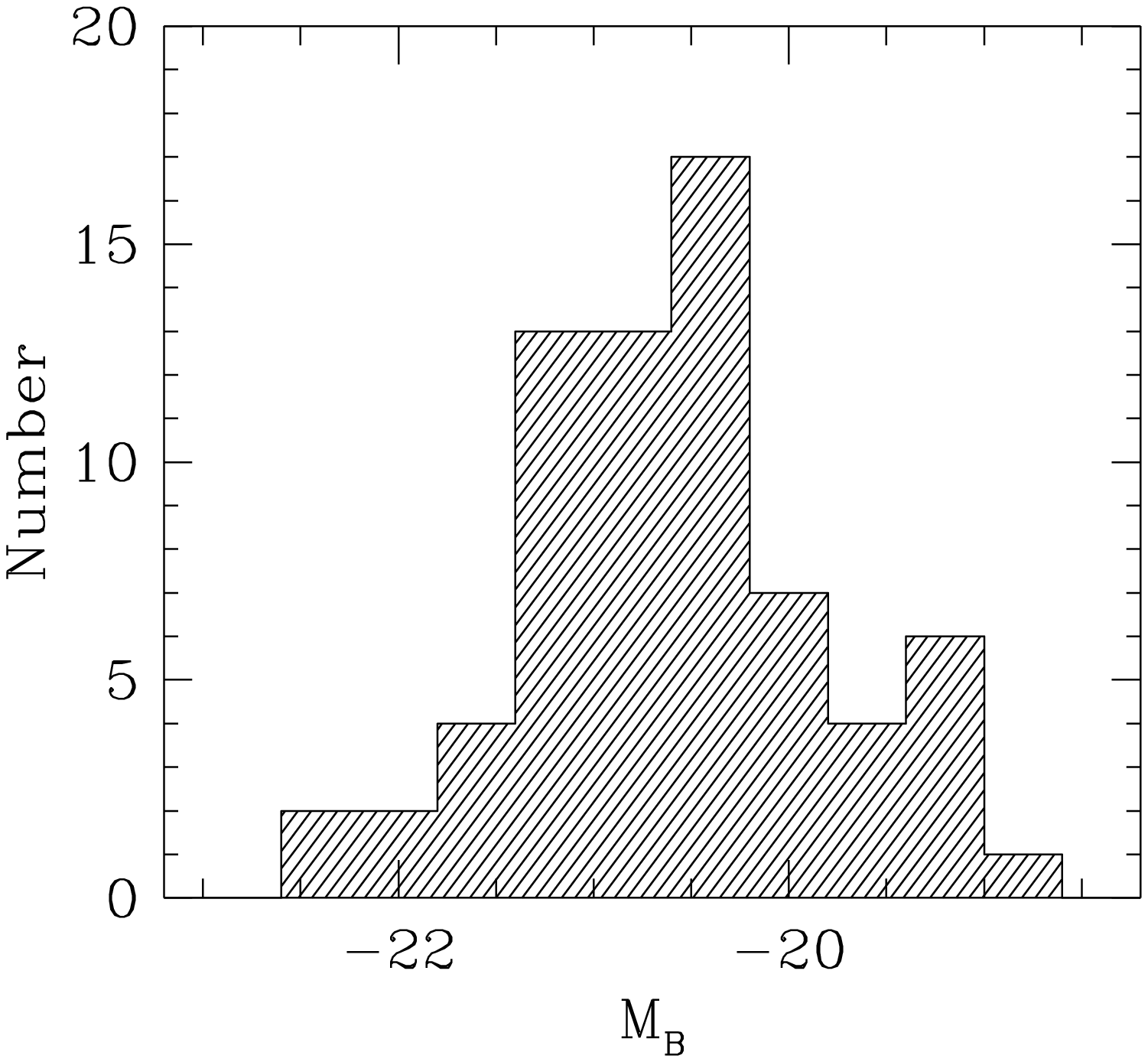}
\caption{}
\end{figure}
\noindent

\begin{figure}
\plotone{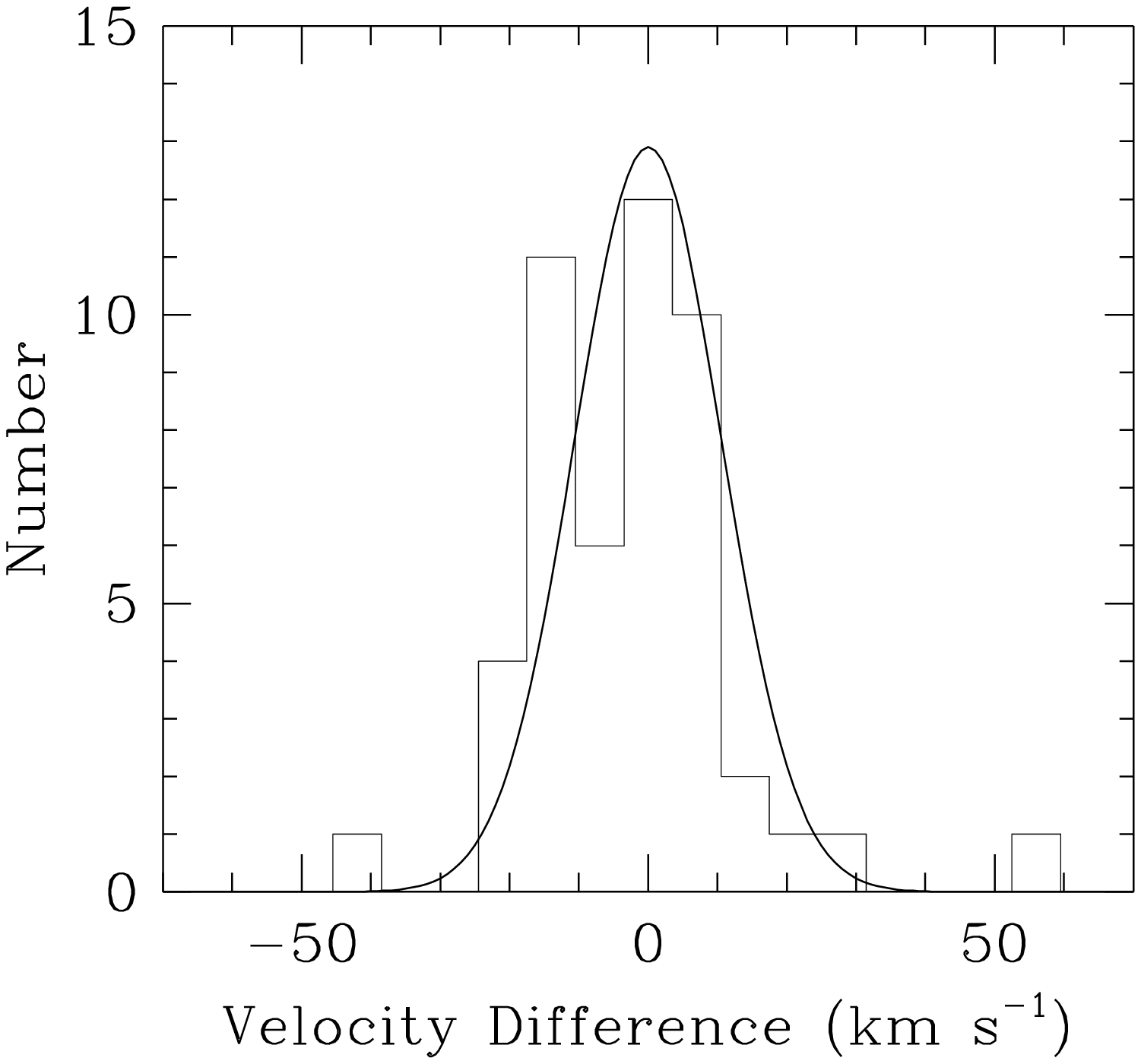}
\caption{}
\end{figure}

\begin{figure}
\plotone{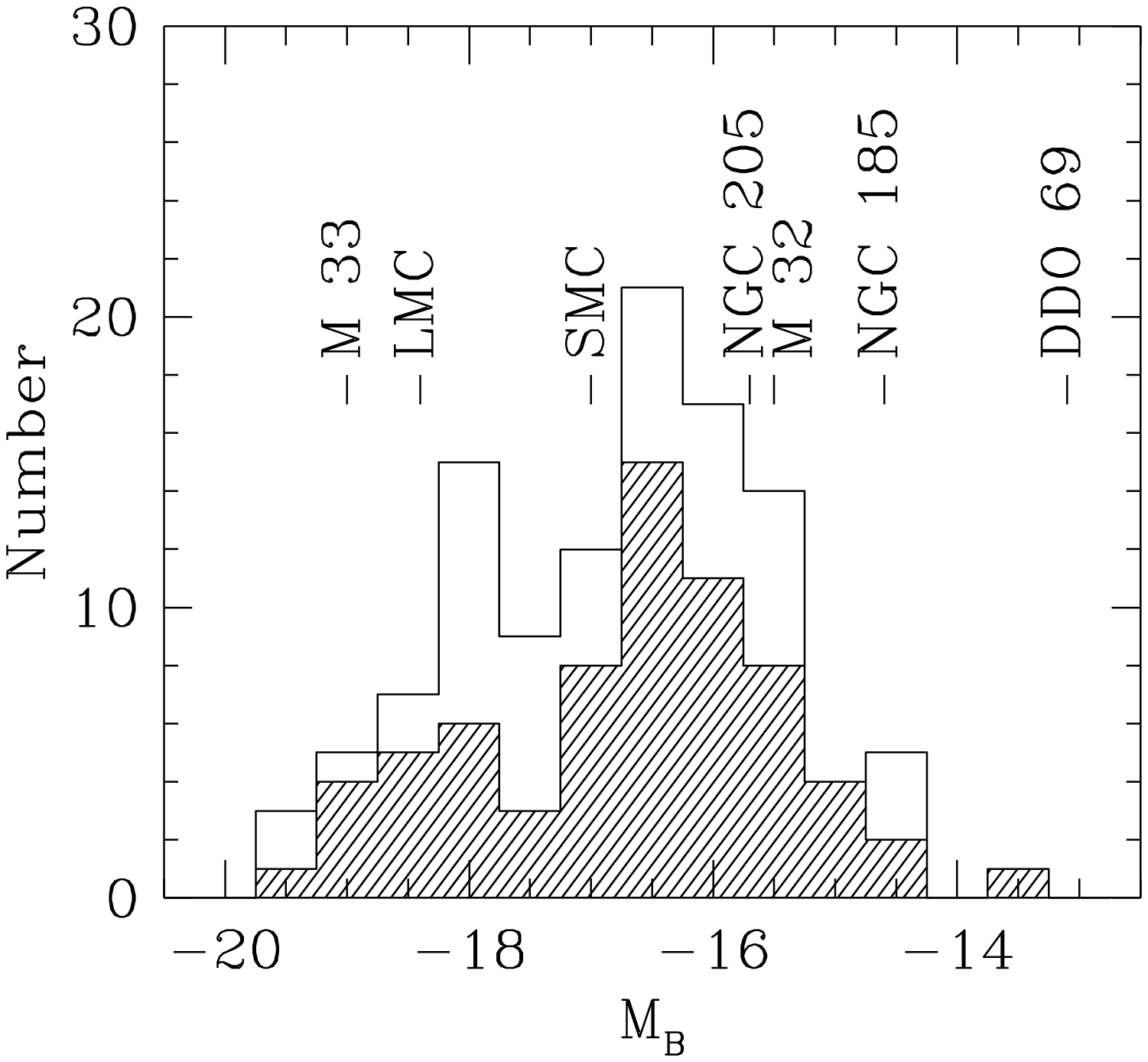}
\caption{}
\end{figure}

\begin{figure}
\plotone{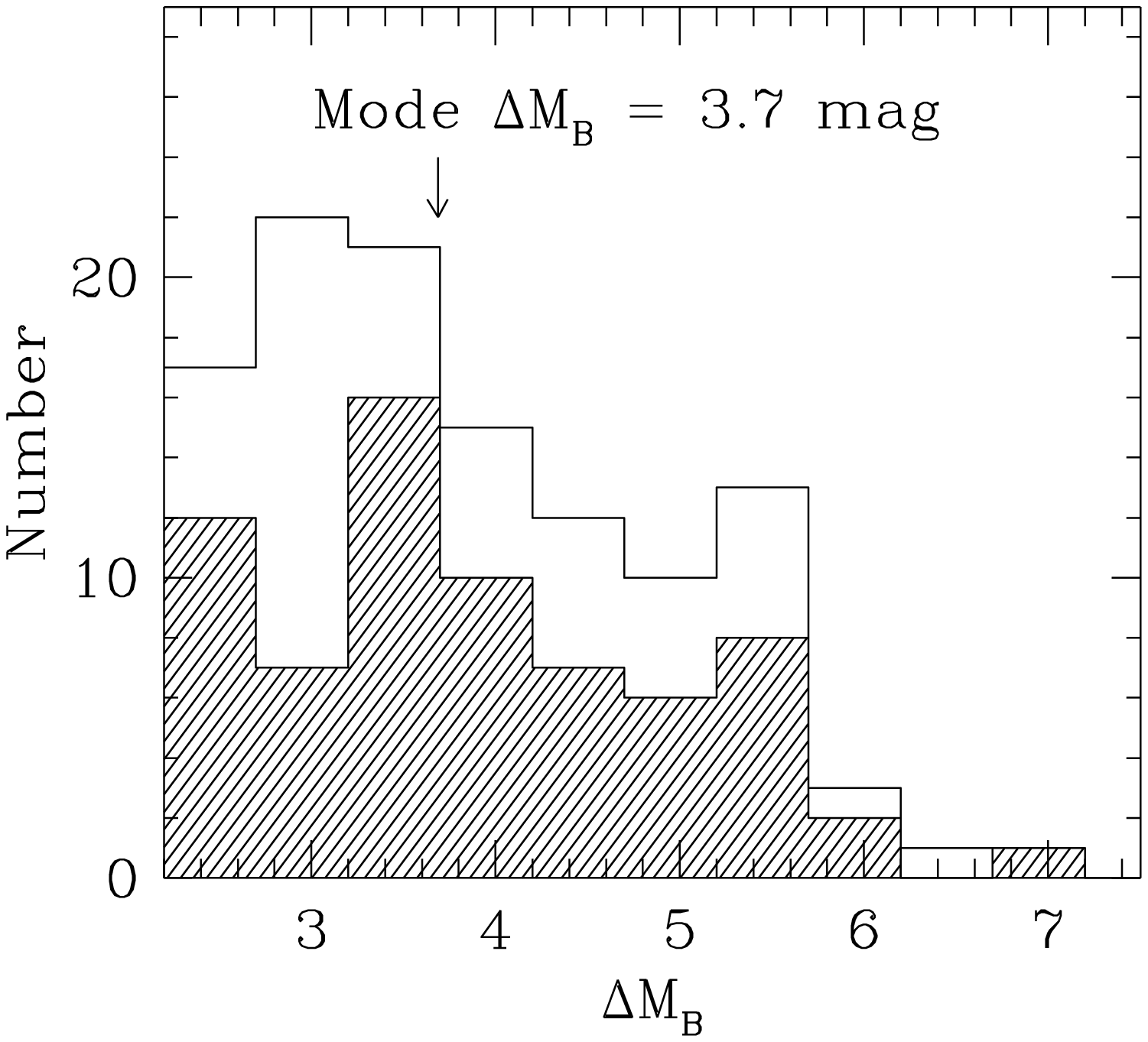}
\caption{}
\end{figure}

\begin{figure}
\plotone{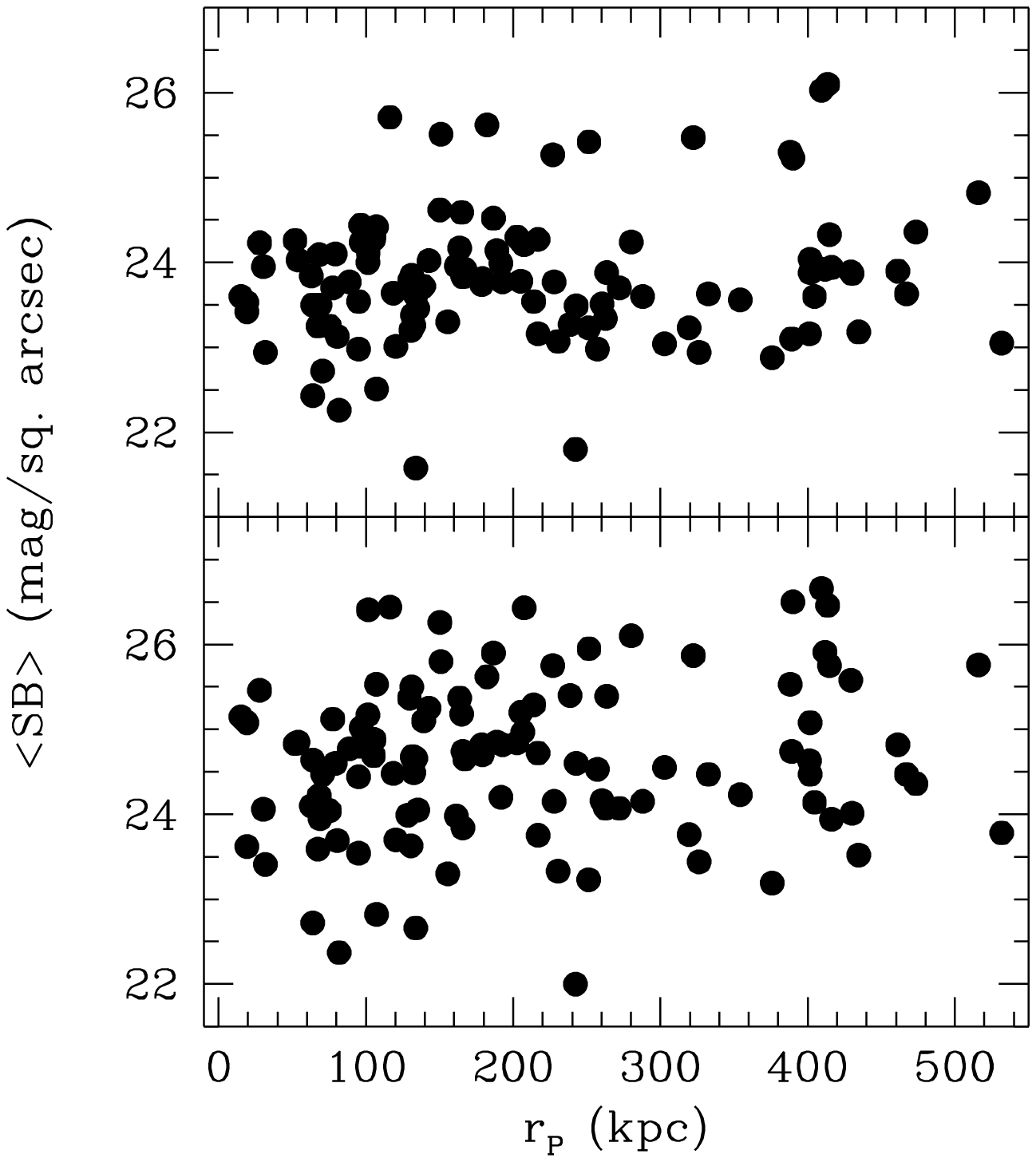}
\caption{}
\end{figure}

\begin{figure}
\plotone{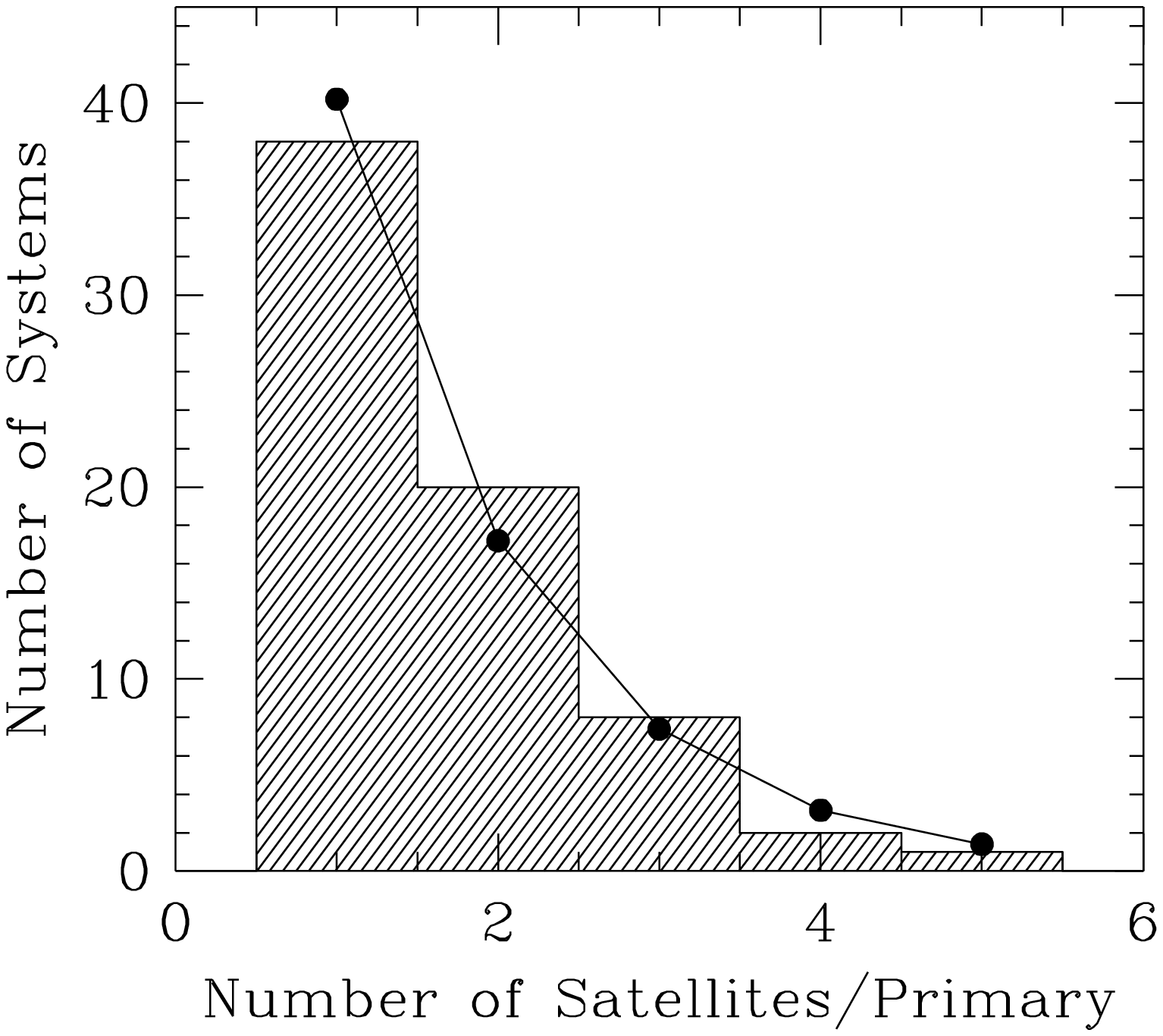}
\caption{}
\end{figure}

\begin{figure}
\plotone{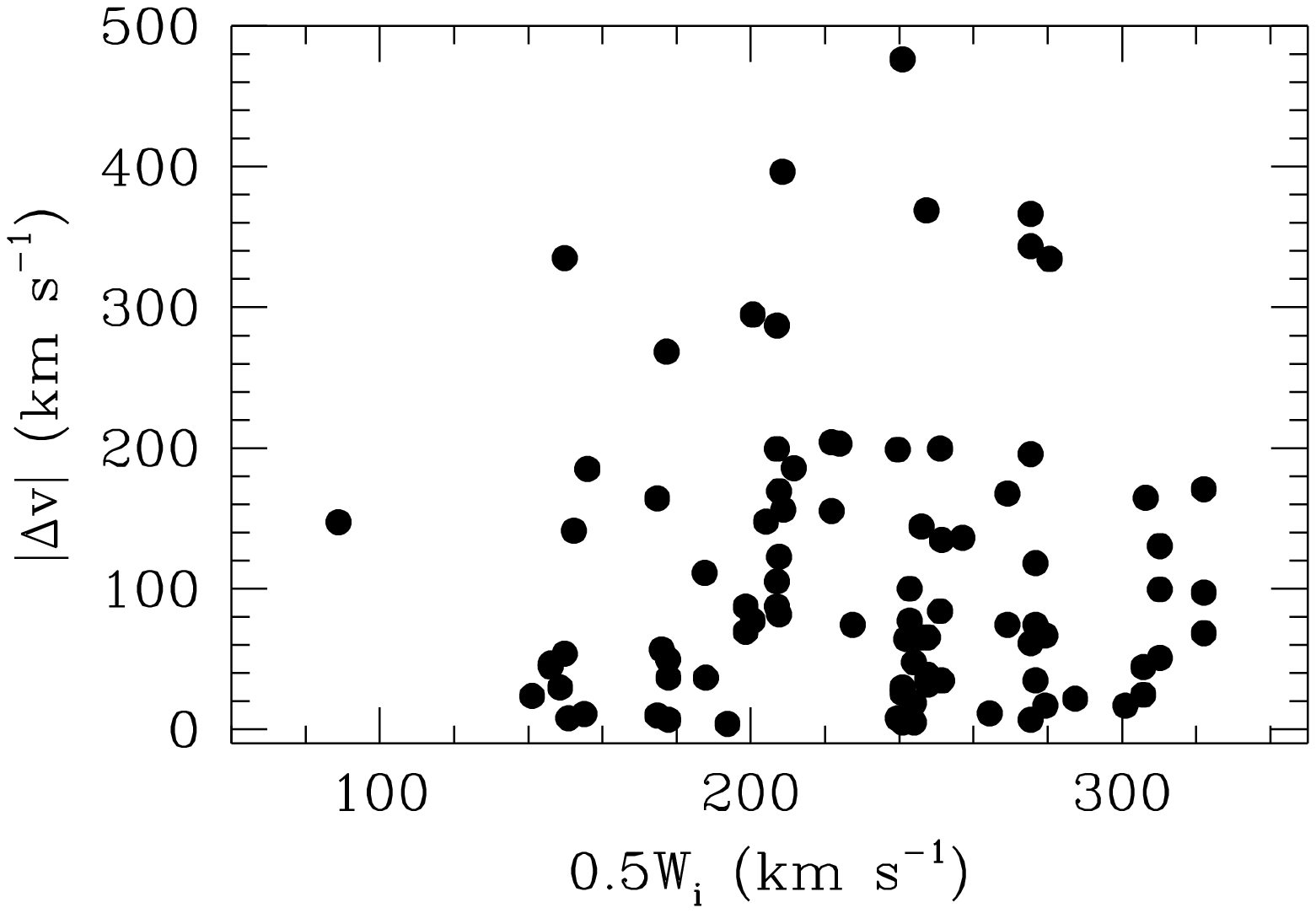}
\caption{}
\end{figure}

\begin{figure}
\plotone{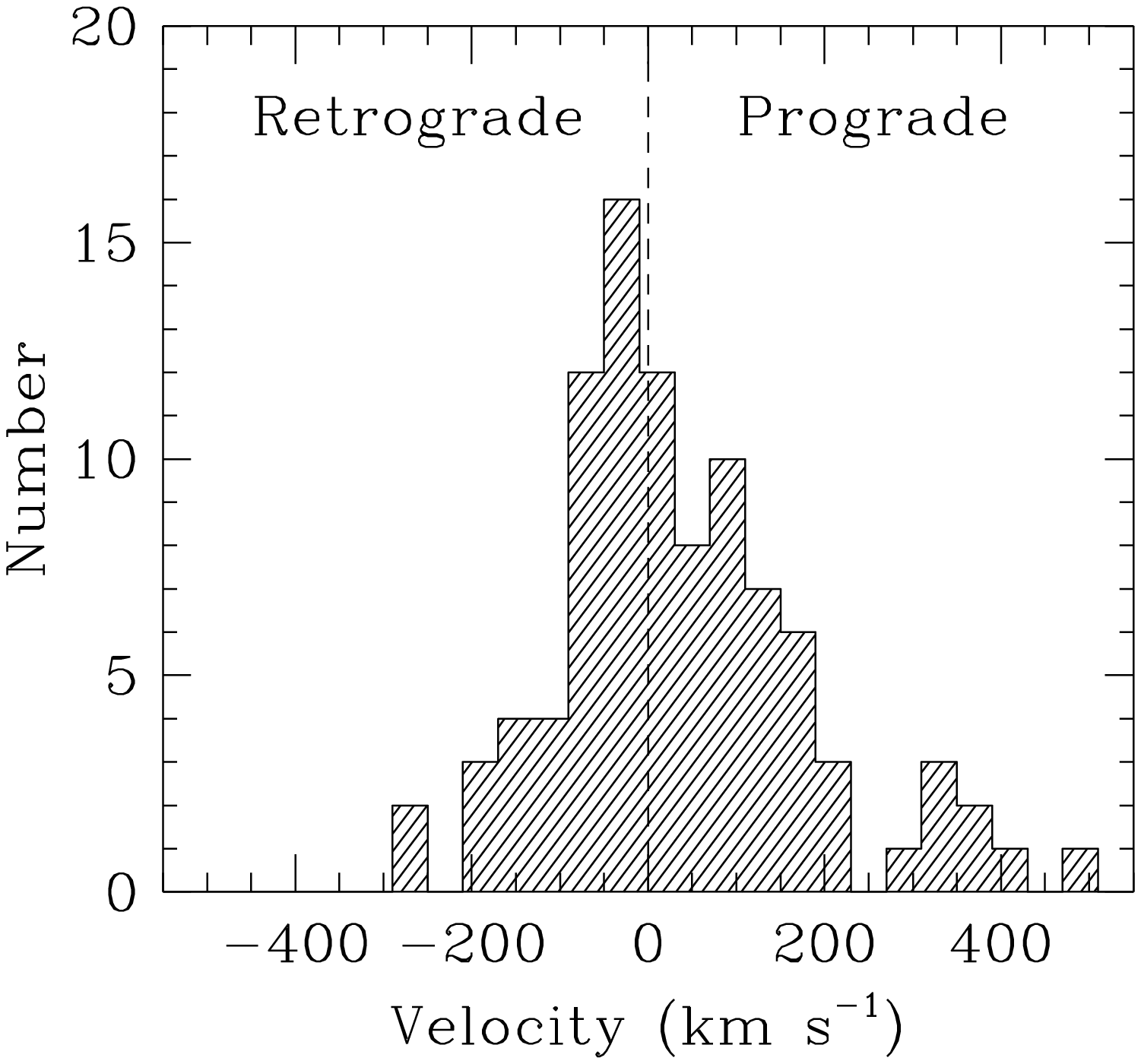}
\caption{}
\end{figure}

\begin{figure}
\plotone{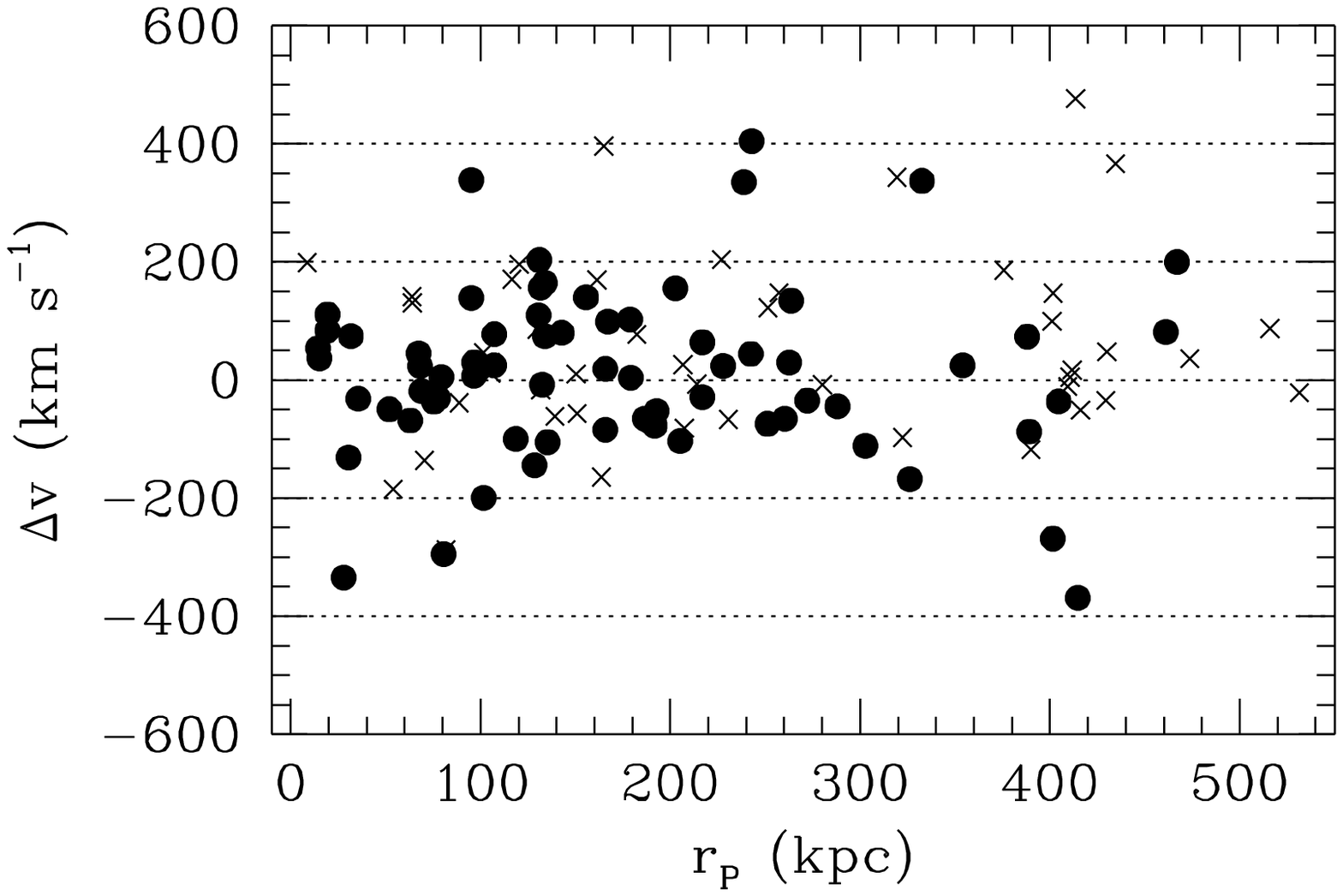}
\caption{}
\end{figure}

\begin{figure}
\plotone{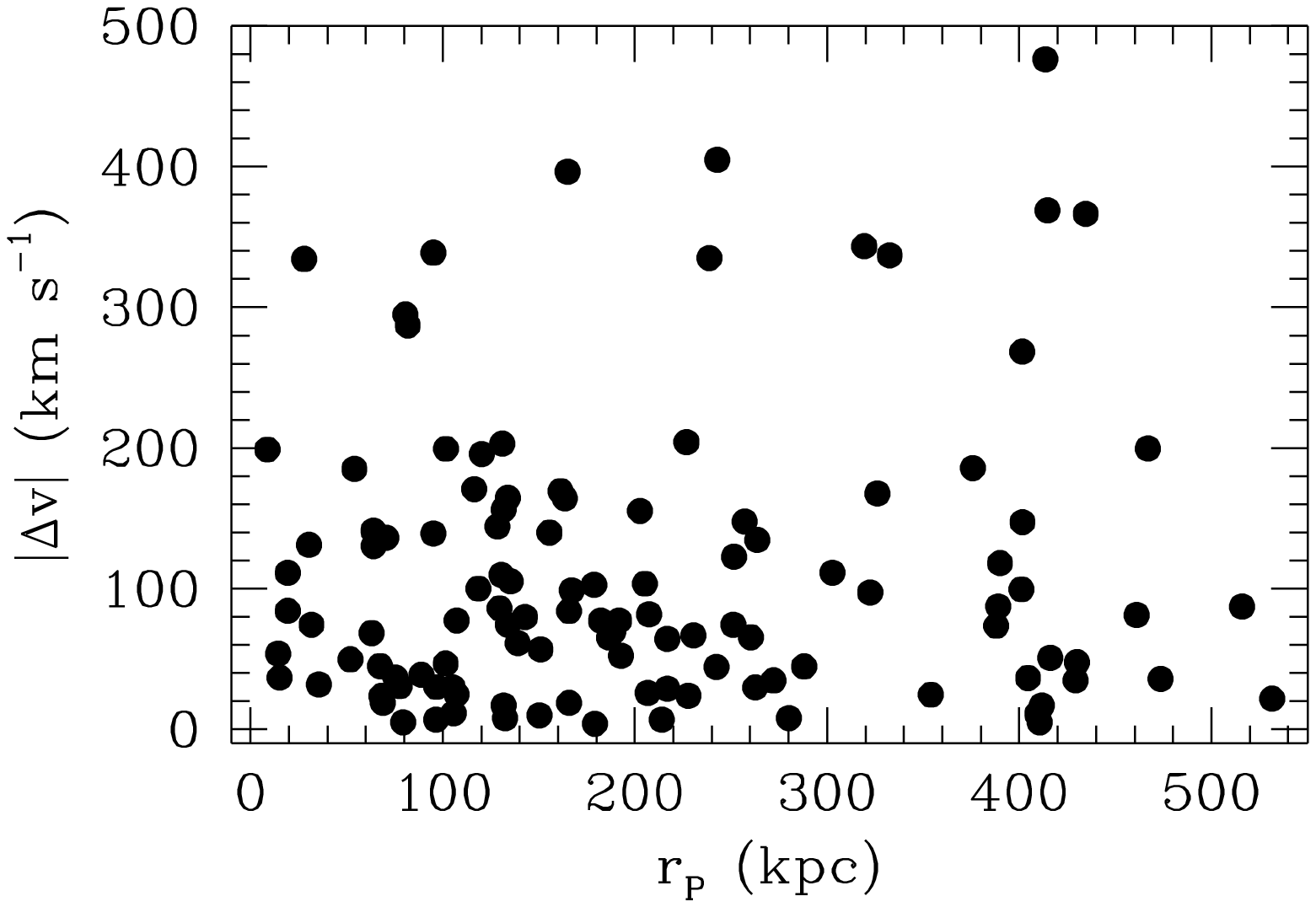}
\caption{}
\end{figure}
\end{document}

%% file: macro.tex
%input macro
%at the start of a tex file. Note that different printers may or
%may not automatically include an offset---if the text is off to
%one side adjust the commands \hoffset and \voffset by adding or
%removing a comment sign %.
\def\chaphead{}

\def\hut{Hubble type\ }
\def\vc{V$_{\rm C}$\ }
\def\mb{M$_{\rm B}$\ }
\def\av{A$_{\rm V}$\ }
\def\lamlam{$\lambda\lambda$}

\def\deg{$^\circ$}
\def\degrees{$^\circ$}
\def\Vlasov{collisionless Boltzmann\ }
\def\lsls{\ll}
\def\grgr{\gg}
\def\erf{\mathop{\rm erf}\nolimits} %error function
\def\eqv{\equiv}
\def\real{\Re e}
\def\imag{\Im m}
\def\ctrline#1{\centerline{#1}}
\def\spose#1{\hbox to 0pt{#1\hss}}
     
\def\={\overline}
\def\sections{\S}
\newcount\notenumber
\notenumber=1
\newcount\eqnumber
\eqnumber=1
\newcount\fignumber
\fignumber=1
\newbox\abstr
\newbox\figca     
\def\yyskip{\penalty-100\vskip6pt plus6pt minus4pt}
     
%\numberpara produces numbered paragraphs with extra space and no indentation
\def\numberpara{\yyskip\noindent}
     
\def\km{{\rm\,km}}
\def\kms{{\rm\ km\ s$^{-1}$}}
\def\kpc{{\rm\,kpc}}
\def\mpc{{\rm\,Mpc}}
\def\etal{{\it et al. }}
\def\eg{{\it e.g., }}
\def\ie{{\it i.e., }}
\def\cf{{\it cf. }}
\def\msun{{\rm\,M_\odot}}
\def\lsun{{\rm\,L_\odot}}
\def\rsun{{\rm\,R_\odot}}
\def\pc{{\rm\,pc}}
\def\cm{{\rm\,cm}}
\def\yr{{\rm\,yr}}
\def\au{{\rm\,AU}}
\def\AU{{\rm\,AU}}
\def\gm{{\rm\,g}}
\def\s{{\rmss}}
\def\dyne{{\rm\,dyne}}
     
%\note macro produces sequentially numbered footnotes at bottom of page
%\foot macro produces sequentially numbered footnotes inserted in text
\def\note#1{\footnote{$^{\the\notenumber}$}{#1}\global\advance\notenumber by 1}
\def\foot#1{\raise3pt\hbox{\eightrm \the\notenumber}
     \hfil\par\vskip3pt\hrule\vskip6pt
     \noindent\raise3pt\hbox{\eightrm \the\notenumber}
     #1\par\vskip6pt\hrule\vskip3pt\noindent\global\advance\notenumber by 1}
\def\propo{\propto}
\def\larrow{\leftarrow}
\def\rarrow{\rightarrow}
\def\sectionhead#1{\penalty-200\vskip24pt plus12pt minus6pt
        \centerline{\bbrm#1}\vskip6pt}
     
%\Dt and \dt put Newton's notation dots above upper and lower case chars
\def\Dt{\spose{\raise 1.5ex\hbox{\hskip3pt$\mathchar"201$}}}    % upper case
\def\dt{\spose{\raise 1.0ex\hbox{\hskip2pt$\mathchar"201$}}}    % lower case
\def\llangle{\langle\langle}
\def\rrangle{\rangle\rangle}
\def\ldotss{\ldots}
\def\del{\b\nabla}
     
% equation numbering
%\new macro produces sequentially numbered equations by writing \eqno(\new)
%at end of displayed equations
\def\new{{\rm\chaphead\the\eqnumber}\global\advance\eqnumber by 1}
%to refer to an equation which is 5 equations back, write "equation (\ref5)"
\def\ref#1{\advance\eqnumber by -#1 \chaphead\the\eqnumber
     \advance\eqnumber by #1 }
%\last macro is like \new except counter is not advanced. Useful for equations
%which are in parts a and b.
\def\last{\advance\eqnumber by -1 {\rm\chaphead\the\eqnumber}\advance
     \eqnumber by 1}
%to name an equation, place command "\eqnam{\Poisson}" before equation, and
%thereafter "equation(\Poisson)" will generate the proper equation number.
\def\eqnam#1{\xdef#1{\chaphead\the\eqnumber}}
     
%figure numbering
%\nfig macro assigns number to a figure
\def\nfig{\chaphead\the\fignumber\global\advance\fignumber by 1}
%\nfiga permits a,b,c etc. to be added to figure number
\def\nfiga#1{\chaphead\the\fignumber{#1}\global\advance\fignumber by 1}
\def\rfig#1{\advance\fignumber by -#1 \chaphead\the\fignumber
     \advance\fignumber by #1}
%\def\fignam#1{\xdef#1{\chaphead\the\fignumber}}
%reference macros. To generate reference to a paper in Ap.J. volume 300, p.123
%write \apj{Claus, S. 1990.}{300}{123}
\def\refindent{\par\noindent\parskip=4pt\hangindent=3pc\hangafter=1 }

\def\apj#1#2#3{\refindent#1,  {ApJ,\ }{#2}, #3}
\def\apjsup#1#2#3{\refindent#1,  {ApJS\ }{#2}, #3}
\def\aasup#1#2#3{\refindent#1,  { AA Sup.,\ }{#2}, #3}
\def\aas#1#2#3{\refindent#1,  { Bull. Am. Astr. Soc.,\ }{#2}, #3}
\def\apjlett#1#2#3{\refindent#1,  { ApJL,\  }{#2}, #3}
\def\mn#1#2#3{\refindent#1,  { MNRAS,\ }{#2}, #3}
\def\mnras#1#2#3{\refindent#1,  { M.N.R.A.S., }{#2}, #3}
\def\annrev#1#2#3{\refindent#1, { ARA \& A,\ }
{\bf2}, #3}
\def\aj#1#2#3{\refindent#1,  { AJ,\  }{#2}, #3}
\def\phrev#1#2#3{\refindent#1, { Phys. Rev.,}{#2}, #3}
\def\aa#1#2#3{\refindent#1,  { AA,\ }{#2}, #3}
\def\nature#1#2#3{\refindent#1,  { Nature,\ }{#2}, #3}
\def\icarus#1#2#3{\refindent#1,  { Icarus, }{#2}, #3}
\def\pasp#1#2#3{\refindent#1,  { PASP,\ }{#2}, #3}
\def\appopt#1#2#3{\refindent#1,  { App. Optics,\  }{#2}, #3}
\def\spie#1#2#3{\refindent#1,  { Proc. of SPIE,\  }{#2}, #3}
\def\opteng#1#2#3{\refindent#1,  { Opt. Eng.,\  }{#2}, #3}
\def\refpaper#1#2#3#4{\refindent#1,  { #2 }{#3}, #4}
\def\refbook#1{\refindent#1}
\def\science#1#2#3{\refindent#1, { Science, }{#2}, #3}
     
\def\chapbegin#1#2{\eject\vskip36pt\par\noindent{\chapheadfont#1\hskip30pt
     #2}\vskip36pt}
\def\sectionbegin#1{\vskip30pt\par\noindent{\bf#1}\par\vskip15pt}
\def\subsectionbegin#1{\vskip20pt\par\noindent{\bf#1}\par\vskip12pt}
\def\topic#1{\vskip5pt\par\noindent{\topicfont#1}\ \ \ \ \ }
     
%\ltsim and \gtsim produce > and < signs with twiddle underneath
\def\ltsim{\mathrel{\spose{\lower 3pt\hbox{$\mathchar"218$}}
     \raise 2.0pt\hbox{$\mathchar"13C$}}}
\def\gtsim{\mathrel{\spose{\lower 3pt\hbox{$\mathchar"218$}}
     \raise 2.0pt\hbox{$\mathchar"13E$}}}
     
%\sec produces arcsec symbol so that 3\sec5 produces 3."5 with the second
%symbol and the period aligned.
\def\sec{\hbox{$^s$\hskip-3pt .}}
\def\gg{\hbox{$>$\hskip-4pt $>$}}
%\hoffset=1.0truein
%\voffset=0.8truein
\parskip=3pt
\def\gapprox{$_ >\atop{^\sim}$}     %Greater than over approximately (wiggle).%
\def\lapprox{$_ <\atop{^\sim}$}     %Less than over approximately.%
\def\apequal{\mathrel{\spose{\lower 1pt\hbox{$\mathchar"218$}}
     \raise 2.0pt\hbox{$\mathchar"218$}}}
\newbox\grsign \setbox\grsign=\hbox{$>$} \newdimen\grdimen \grdimen=\ht\grsign
\newbox\simlessbox \newbox\simgreatbox
\setbox\simgreatbox=\hbox{\raise.5ex\hbox{$>$}\llap
     {\lower.5ex\hbox{$\sim$}}}\ht1=\grdimen\dp1=0pt
\setbox\simlessbox=\hbox{\raise.5ex\hbox{$<$}\llap
     {\lower.5ex\hbox{$\sim$}}}\ht2=\grdimen\dp2=0pt
\def\gtorder{\mathrel{\copy\simgreatbox}}
\def\ltorder{\mathrel{\copy\simlessbox}}
\def\simgreat{\mathrel{\copy\simgreatbox}}
\def\simless{\mathrel{\copy\simlessbox}}

%% file: table1.tex
\halign to \hsize{#\hfil\quad&#\hfil\cr
\noalign{\bigskip}
\noalign{\hrule \vskip 2pt \hrule \medskip}
N259a&At high redshift (45,400 km s$^{-1}$) from follow-up spectroscopy \cr
N1459 Field&Galaxy at r$_p = 792$ kpc with $\Delta m_B = 0.2$\cr
N4679 Field&Galaxy at r$_p = 120$ kpc with $\Delta m_B = 1.3$\cr
N5073 Field&N5073b has  $\Delta m_B = 1.5$ \cr
N5085 Field&N5085a superposed on a bright star (current $\Delta m_B$
estimate is 1 mag)\cr
N5254 Field&N5254a has $\Delta m_B = 1.5$ \cr
N5297 Field&N5297a has $\Delta m_B = 1.5$\cr
N5768 Field&N5768a has $\Delta m_B = 1.1$\cr
N6943b&CTIO velocity was not confirmed with follow-up spectroscopy \cr
\noalign{\vskip -8pt}
&(no reliable velocity measured)\cr
A2120 Field&Galaxy at r$_p = 435$ kpc with $\Delta m_B = 1.7$\cr
N7083 Field&Galaxy at r$_p = 409$ kpc with $\Delta m_B = 1.0$\cr
N7716 Field&N7716a has $\Delta m_B = 1.5$\cr
\noalign{\smallskip \hrule}
}